\documentclass[conference]{IEEEtran}
\IEEEoverridecommandlockouts
\usepackage{cite}
\usepackage{amsmath,amssymb,amsfonts}
\usepackage{algorithmic}
\usepackage{graphicx}
\usepackage{textcomp}
\usepackage{xcolor}

\usepackage{graphicx}
\usepackage{longtable}
\usepackage{booktabs}

\def\BibTeX{{\rm B\kern-.05em{\sc i\kern-.025em b}\kern-.08em
    T\kern-.1667em\lower.7ex\hbox{E}\kern-.125emX}}
\begin{document}

\title{Time and Frequency Domain-based Anomaly Detection in Smart Meter Data for Distribution Network Studies

\thanks{This work has been in part supported by the the HEDGE-IoT project. The project received funding from the European Union’s Horizon Europe research and innovation programme under the grant agreement Nº 101136216. Views and opinions expressed are however those of the author(s) only and do not necessarily reflect those of the European Union or European Climate, Infrastructure and Environment Executive Agency. Neither the European Union nor the granting authority can be held responsible for them.

Employment of Petar Labura is fully funded by Croatian Science Foundation (HRZZ) within programme NPOO-DOK-2023-10.}
}

\author{\IEEEauthorblockN{Petar Labura}
\IEEEauthorblockA{\textit{University of Zagreb} \\
\textit{Faculty of Electrical Engineering}\\
\textit{and Computing}\\
Zagreb, Croatia \\
petar.labura@fer.unizg.hr}
\and
\IEEEauthorblockN{Tomislav Antić}
\IEEEauthorblockA{\textit{University of Zagreb} \\
\textit{Faculty of Electrical Engineering}\\
\textit{and Computing}\\
Zagreb, Croatia \\
tomislav.antic@fer.unizg.hr}
\and
\IEEEauthorblockN{Tomislav Capuder}
\IEEEauthorblockA{\textit{University of Zagreb} \\
\textit{Faculty of Electrical Engineering}\\
\textit{and Computing}\\
Zagreb, Croatia \\
tomislav.capuder@fer.unizg.hr}
}

\maketitle

\begin{abstract}
The widespread integration of new technologies in low-voltage distribution networks on the consumer side creates the need for distribution system operators to perform advanced real-time calculations to estimate network conditions. In recent years, data-driven models based on machine learning and big data analysis have emerged for calculation purposes, leveraging the information available in large datasets obtained from smart meters and other advanced measurement infrastructure. 
However, existing data-driven algorithms do not take into account the quality of data collected from smart meters. They lack built-in anomaly detection mechanisms and fail to differentiate anomalies based on whether the value or context of anomalous data instances deviates from the norm.
This paper focuses on methods for detecting and mitigating the impact of anomalies on the consumption of active and reactive power datasets. It proposes an anomaly detection framework based on the Isolation Forest machine learning algorithm and Fast Fourier Transform filtering that works in both the time and frequency domain and is unaffected by point anomalies or contextual anomalies of the power consumption data. The importance of integrating anomaly detection methods is demonstrated in the analysis important for distribution networks with a high share of smart meters.

\end{abstract}


\section{Introduction} \label{intro}

In order to stimulate and accelerate the integration of new technologies into low-voltage (LV) networks, while simultaneously ensuring the safety and reliability of energy supply in distribution systems, distribution system operators (DSOs) are required to perform advanced calculations in near real-time with high accuracy.

For this reason, an increasing amount of research is directed toward implementing data-driven solutions and utilizing the information within large datasets of measurement data obtained from smart meters. Distribution networks were traditionally characterized by low observability and the exact state of the network was unknown due to the low share of advanced metering infrastructure (AMI). However, energy transition and changes in distribution networks created the need for installing smart meters in distribution network nodes. Integration of smart meters creates numerous opportunities and allows DSOs to better plan and operate the network \cite{8322199}. There is a broad spectrum of possible applications of data-driven methods. Models like physics-informed neural networks (PINNs) are used to capture complex network relationships using deep neural networks that incorporate basic physical laws into the loss function used for training, therefore achieving faster convergence and improved prediction capabilities \cite{9115822}. PINNs have been explored for power system stability assessment \cite{9813554, 10005095}, as well as for voltage calculations \cite{10517953} and state estimation \cite{10167837}. 

In order to completely overcome the problem of unknown topological connectivity for the purposes of voltage calculation, model-free voltage calculation algorithms are proposed \cite{9975837}. These algorithms use neural networks to model complex non-linear relationships between measured datasets, thus bypassing the need for any information on electrical network models. 
Machine learning methods have also been used for load forecasting purposes, utilizing reinforcement Q-learning \cite{8813103} or deep learning techniques such as long-short term memory neural networks \cite{9528900}. Data-driven prediction algorithms are also seeing increasing implementation for photovoltaic (PV) generation forecasting. In \cite{9399287}, a sky image prediction model is developed for accurate short-term PV generation predictions.

In general, the main drawback of data-driven methods is their dependency on the quality of gathered data. In most cases, however, studies do not account for potentially anomalous or noisy data gathered by smart meters. For purposes of power flow calculations, outlier-immune methods have been discussed in \cite{10669038}. Authors of \cite{8809905} explored probabilistic autoencoders for outlier detection of power system measurements and their reconstruction. 
The work in \cite{9760038} is focused on sequential anomaly detection for monitoring purposes by examining continuous phasor measurements.
A more advanced anomaly detection solution for purposes of state estimation and energy management systems for power grids is presented in \cite{ASEFI2023101116}. This paper investigates more complex models to detect and identify the origin of anomalies that bypass the $\chi^2$-test, combining analytical and machine learning methods. That being said, anomaly detection has rarely been analyzed in great detail when developing machine learning-based solutions concerning state estimation, voltage calculations, load forecasting, or PV generation predictions. In many cases, data is analyzed as it is collected without verifying the accuracy, although this can directly impact the calculations, as well as subsequent decision-making. If anomaly detection is considered, usually a simplistic anomaly detection model is used without a detailed explanation regarding which specific type of anomalies is considered. A more detailed anomaly detection study evaluating machine learning methods is performed in \cite{9726670}, however, it does not examine the direct source of anomalies. Our work provides a detailed framework that can detect anomalies regardless of their category: This paper explores more complex and unconventional techniques for anomaly detection, such as isolation forest machine learning method \cite{isolation_forest} and Fast Fourier Transform filter. 

There is a multitude of potential causes of registered anomalies in smart meter datasets, such as sensor malfunction, calibration issues, power outages, sampling rate errors,  thermal noise, electromagnetic interference, communication network malfunction, lightning strikes, switching operations, or cybersecurity threats. Upon a detailed literature review, we conclude that the main gap in the literature is not considering data quality and the presence of anomalies in the data used for training, which is a standard event in real-world data obtained from smart meter measurements. Furthermore, when anomaly detection methods are discussed, it is very rarely justified why specific methods are used as opposed to alternatives. We define an anomaly detection framework that can be used in various distribution system studies, in both the time and frequency domain. To deal with the identified research gaps, the following contributions concerning anomaly detection are proposed:
\begin{itemize}
    \item Elimination of consumption data point anomalies in the time domain using the Isolation Forest machine learning algorithm.
    \item Fast Fourier Transform filter for smoothing the data in frequency domain and significantly reducing the influence of contextual anomalies in consumption data. 
    \item Investigating the accuracy of different anomaly detection methods on a smart meter dataset and validating its importance on a network analysis relevant to a DSO.
\end{itemize}

The remainder of this paper is structured in a following way: Section \ref{methods} describes the mathematical formulation of algorithms used for anomaly detection. Section \ref{casestudy} divides the research into 2 case studies, depending on the examined type of anomalies. Section \ref{results} applies the methodology principles to practical data and in Section \ref{discussion} these anomaly detection results are presented in practical use on an example of data-driven voltage predictions. Final conclusions are drawn in Section \ref{conclusion}.

\section{Methodology} \label{methods}

Data-driven solutions based on statistical and machine learning methods are witnessing increased application in distribution networks. The proliferation of smart meter installations has made large sets of measurement data available for distribution system operators (DSOs) to analyze. However, the majority of advanced network analyses algorithms assume idealistic data quality, without considering the sensitivity to anomalies and outliers present in the dataset. Upon a thorough examination of the existing literature, we identify the lack of proper investigation of the impact of data quality in data-driven network analyses. In our work, we put the focus on researching different anomaly detection methods and their correlation with the accuracy in estimating voltage values in an LV network.

Anomaly detection generally refers to the process of identifying rare data instances or unusual events that significantly deviate from the usual pattern or behavior. In that sense, we differentiate the term outlier from the term anomaly. An outlier is a statistical term, whilst an anomaly encompasses a broader concept, taking into account the context in which data aberration occurred. An outlier is defined as a subset of anomalies, in the sense that all outliers are anomalies, but not all anomalies are necessarily outliers. Outliers are generally easier to detect, using conventional statistical methods such as Interquartile Range (IQR) or Mahalanobis Distance. This work identifies and focuses on two distinct types of anomalies:
\begin{itemize}
  \item \textbf{Point anomalies} -- singular data points that deviate from the norm.
  \item \textbf{Contextual anomalies} -- data points that are normal in some contexts but anomalies in a different context.
\end{itemize}

This paper examines exclusively unsupervised anomaly detection techniques that assume data is unlabeled. Due to anomalous data naturally being difficult to recognize, it is not expected that a labeled dataset could be provided for the classification of anomalies. 

The data for the observed network refers to the real suburban LV network, where historical datasets for active power \textbf{P}, reactive power \textbf{Q}, and voltage magnitude \textbf{V} measurements have been collected into a database. 

\subsection{Point anomalies}

Point anomalies are simulated as a stochastic process and added to a copy of the original database to simulate the presence of anomalous data. For this purpose, a stochastic process representing point anomalies influenced by a Bernoulli random variable is simulated as in \eqref{eq:impulse_noise}. $P(t)$ represents a point anomaly-imputing random process that can obtain only two values. A spike value of intensity $L$ for situations when the Bernoulli process is realized with probability $p$ and a value of zero otherwise. Generally, $p$ is a small value since point anomalies occur infrequently. This random process is then added to dataset measurements to simulate the influence of point anomalies.

\begin{equation}
P(t \,|\, L, p) = \begin{cases} 
L & \text{with probability } p \\
0 & \text{with probability } (1 - p)
\end{cases}
\label{eq:impulse_noise}
\end{equation}

With the purpose of filtering point anomalies, an Isolation Forest machine learning algorithm is applied. Isolation forest is a tree-based ML algorithm similar to conventional random forest methods but developed solely for the purpose of anomaly detection. The core goal of isolation forests is isolating anomalous data instances, as opposed to developing a model around them. It is assumed that anomalies are rare and substantially different than normal data points, thus being easier to isolate.  

The algorithm works by building a binary tree structure. It starts by choosing a random feature and then a random split point within the range of values of that feature. This partition process is recursive and continues until the remainder of the tree is built and every single data point is independently isolated. Upon completion of this tree structure, a path length metric is defined as the number of partitions taken to isolate a data point. Anomalies, being a minority of total data points, are expected to be easier to isolate, due to having a shorter average path length than normal data points. The shorter path length in a tree increases the probability that a selected data instance is an anomaly. However, in order to increase the overall accuracy of the model, many isolation trees are aggregated to produce a more stable result that is less sensitive to randomness.  

Subsequently, an anomaly score $s(x; n)$ is defined for each data point $x$ in a sample size of $n$. The anomaly score is calculated considering the average path length of aggregated isolation trees according to \eqref{eq:anomaly_score}, where $E(h(x))$ represents the expected average path length of all isolation trees and $c(n)$ represents a normalization constant. 

\begin{equation}
    s(x; n) = 2^{-\frac{E(h(x))}{c(n)}}
    \label{eq:anomaly_score}
\end{equation}

Lastly, a contamination rate hyperparameter $\alpha$ is specified. Contamination rate refers to the percentage of data expected to be anomalies. Data instances with $\alpha$ lowest anomaly scores are then labeled as anomalies and replaced with a more representative value, such as the median of that feature.

\subsection{Contextual anomalies}

Contextual anomalies are defined as data values anomalous when they occur in some context, but not in another context. This differentiates contextual anomalies from the usual point anomalies and also makes them more difficult to detect. The reason why contextual anomalies are more difficult to detect than point anomalies is their dependency on context-specific features (i.e. time, location)  rather than just their values. For example, point anomalies may be detected by algorithms that check if a value is extreme but in the case of contextual anomalies, the value of the anomalous data instance might be normal, thus requiring the detection algorithm to depend on surrounding data. 

Contextual anomalies are simulated similarly to point anomalies but with a difference given in \eqref{eq:context_anomalies}. $C(t)$ represents the contextual anomaly imputing random process and $\mu$ represents the mean daily peak value imputed. The core difference lies in the values and occurrence time of chosen instances. Specifically, we have chosen time of occurrence as a context for anomaly simulation and we are imputing mean consumption measurement to randomly chosen instances during night hours (i.e. 11 PM to 7 AM measurements). What this does is simulate a problem of otherwise normal consumption measurements occurring unexpectedly. Consumption values that would be normal during the day are recorded at night hours when consumption is expected to be very low.    

\begin{equation}
C(t \,|\, \mu, p) = \begin{cases} 
\mu & \text{with probability } p \text{, only if $t$ is a night hour}\\
0 & \text{otherwise }
\end{cases}
\label{eq:context_anomalies}
\end{equation}

For purposes of filtering contextual anomalies, a Fast Fourier Transform (FFT) filter is used. FFT filter works by filtering a signal in a frequency domain and smoothing the signal by eliminating specific frequency components. The main idea behind the FFT filter is taking into account the fact that anomalies often introduce high-frequency components to the signal. Sudden spikes that are occurring are usually manifested as high-frequency noise that can be filtered using a low-pass filter. The first step of the FFT filter involves performing a Fourier transform on the anomalous signal $x[n]$ according to \eqref{eq:fft}, where $X[k]$ represents the Fourier transform in a frequency domain, $N$ is the total number of samples, $k$ is a frequency domain index and $n$ is a temporal domain index.

\begin{equation}
    X[k] = \sum_{n=0}^{N-1} x[n] e^{-2\pi i \frac{kn}{N}}
    \label{eq:fft}
\end{equation}

Once the signal is decomposed into its frequency components using FFT, a low-pass frequency filter is applied to selectively attenuate high-frequency components or noise. This produces a smoothing effect thus reducing the influence of high-frequency disturbances such as contextual anomalies, while simultaneously preserving the core characteristics of the signal. After applying a low-pass frequency filter, a smoothed signal is reconstructed by applying an Inverse Fast Fourier Transform (IFFT), but now with only the most relevant components of the spectrum present in the time-domain representation. IFFT formula is given in \eqref{eq:ifft}, where $\hat{x}[n]$ represents the reconstructed signal where the influence of contextual anomalies is reduced.

\begin{equation}
    \hat{x}[n] = \frac{1}{N} \sum_{k=0}^{N-1} X[k] e^{2\pi i \frac{kn}{N}}
    \label{eq:ifft}
\end{equation}

Summarizing, a Fourier Transform filter for contextual anomalies is a 3-step process:
\begin{enumerate}
    \item The FFT of an anomalous signal is obtained.
    \item High-frequency harmonics are eliminated using a low-pass filter.
    \item The IFFT is used to reconstruct the signal.
\end{enumerate}

For validation purposes, if a labeled anomaly dataset is available, then an F1-score metric is used \eqref{eq:f1_score}, because we consider it most balanced and informative, since it takes into account both false positive and false negative values.

\begin{equation}
\begin{aligned}
\text{F1 score} &= 2 \cdot \frac{\text{Precision} \cdot \text{Recall}}{\text{Precision} + \text{Recall}} \\
\text{Precision} &= \frac{TP}{TP + FP}, \quad 
\text{Recall} = \frac{TP}{TP + FN}
\end{aligned}
\label{eq:f1_score}
\end{equation}

\section{Case study} \label{casestudy}

In this section, the case study analysis is outlined. The whole methodology is validated on a real Croatian three-phase LV network with 64 nodes and 43 end-users. In this network, data has been gathered over a period of one year with a 30-minute sampling resolution, making a total of 17,520 data instances per measurement register and a total of more than 2 million data instances analyzed. 

In the first case study, an isolation forest machine learning algorithm is trained for anomaly detection in consumption data. For all detected anomalies, a median value is imputed. The second case study concerns consumption data contextual anomalies over time. Contextual anomalies are usually represented by high-frequency components, so their influence can be significantly smoothed.

In summary, the two case studies explored in this paper are as follows:
\begin{itemize}
    \item \textbf{Case study 1 (CS1):} Isolation forest machine learning algorithm for point anomaly detection of smart meter consumption data.
    \item \textbf{Case study 2 (CS2):} Fourier transform filter to minimize the impact of contextual anomalies by smoothing the consumption data in the frequency domain. 
\end{itemize}

\section{Results} \label{results}

In this section, the results of the analysis are presented following the methodology described in Section \ref{methods}.

\subsection{Case study 1}

\begin{figure}[htbp]
    \centering
    \includegraphics[width=0.48\textwidth]{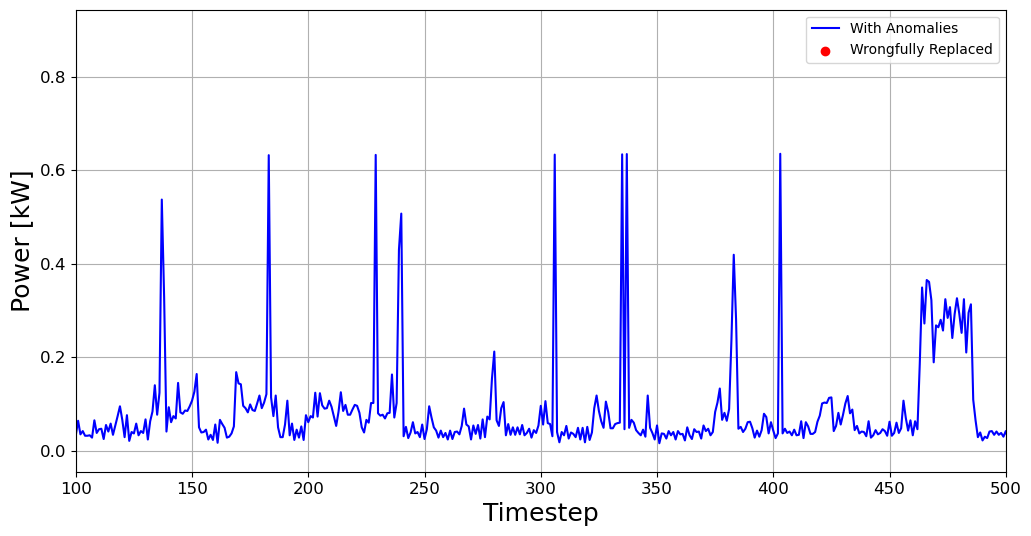} 
    \caption{Data with point anomalies}
    \label{fig:point_anomalies}
\end{figure}

The first case study is concerned with point anomalies. Point anomalies are simulated as outlined in the methodology chapter. A spike impact of 6 standard deviations was chosen, with a $p=2\%$ probability of occurrence. This frequent occurrence of point anomalies represents an extreme case; it is higher than expected in practical applications. However, the machine learning algorithm, Isolation Forest, can handle it effectively. An example of a column with point anomalies added is shown in Fig. \ref{fig:point_anomalies}.

\begin{figure}[htbp]
    \centering
    \includegraphics[width=0.48\textwidth]{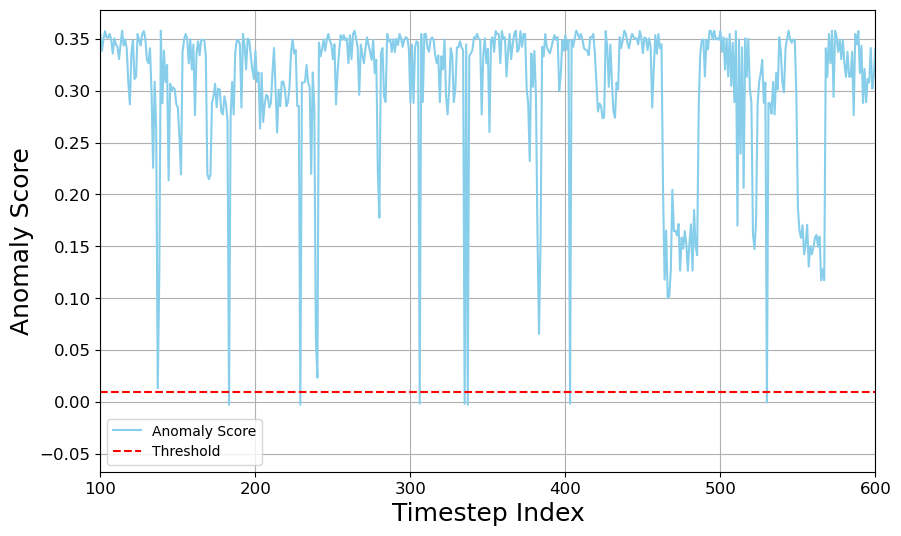} 
    \caption{Anomaly scores with Isolation forest}
    \label{fig:i_forest_anomaly_scores}
\end{figure}

Following this point anomaly imputation, the Isolation Forest algorithm is trained as described in the methodology section. First, isolation trees are created by building binary trees with random feature partitions, making it easier to isolate anomalies. To mitigate the impact of randomness, multiple isolation tree models are aggregated, forming an isolation forest. An anomaly score is then calculated for each data instance by analyzing the aggregated isolation trees. Once the anomaly score is determined, the contamination rate $\alpha$ is chosen as a hyperparameter that regulates the proportion of instances that the algorithm will classify as anomalous. For this case study, a value $\alpha=1\%$ was considered appropriate. Lower anomaly scores indicate a higher probability that a given data point is anomalous. Fig. \ref{fig:i_forest_anomaly_scores} shows a plot of all anomaly scores for an observed LV network node together with a contamination rate threshold. 

\begin{figure}[htbp]
    \centering
    \includegraphics[width=0.48\textwidth]{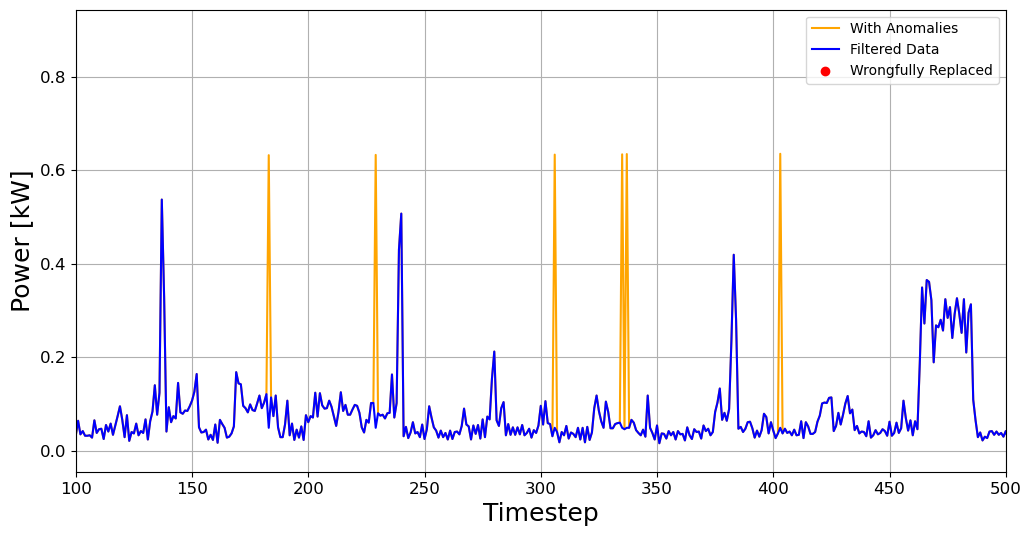} 
    \caption{Anomaly detection with Isolation forest}
    \label{fig:i_forest_anomaly_detection}
\end{figure}

Identified anomalies, with an anomaly score below the contamination rate threshold, are then replaced by the statistical median of the corresponding features. Fig. \ref{fig:i_forest_anomaly_detection} shows the difference between the anomalous consumption measurement time series compared to the one where point anomalies are identified and replaced. 

Although isolation forest anomaly detection is not flawless and may sometimes incorrectly classify certain instances, overall its performance delivers excellent results. F1-score of Isolation forest detection exceeds 77\% which is very satisfying considering the highly unbalanced nature of the anomaly dataset. The Isolation forest algorithm also works very well in conjunction with other data-driven methods.

\subsection{Case study 2}

\begin{figure}[htbp]
    \centering
    \includegraphics[width=0.48\textwidth]{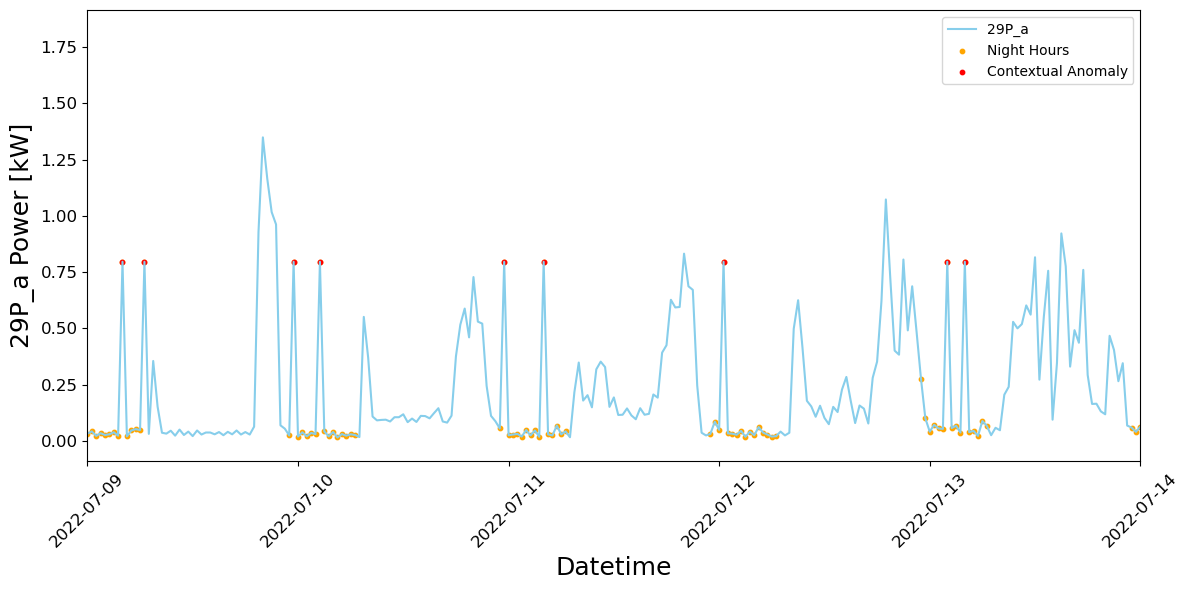} 
    \caption{Contextual anomalies example}
    \label{fig:contextual_anomaleis}
\end{figure}

CS2 examines the impact of contextual anomalies in time. As described in the Section \ref{methods}, a mean daily peak consumption value is added to timesteps corresponding to night hours (11 PM-7 AM) with $p=5\%$. What makes contextual anomaly detection challenging is the fact that registered consumption value is not anomalous in itself, but rather the time of its occurrence is anomalous. As can be seen in in Fig. \ref{fig:contextual_anomaleis}, there is a multitude of non-anomalous consumption values larger than the ones flagged as contextual anomalies. However, these higher consumption values mostly occur during the daytime hours, when end-users consumption is expected to be increased. 

\begin{figure}[htbp]
    \centering
    \includegraphics[width=0.48\textwidth]{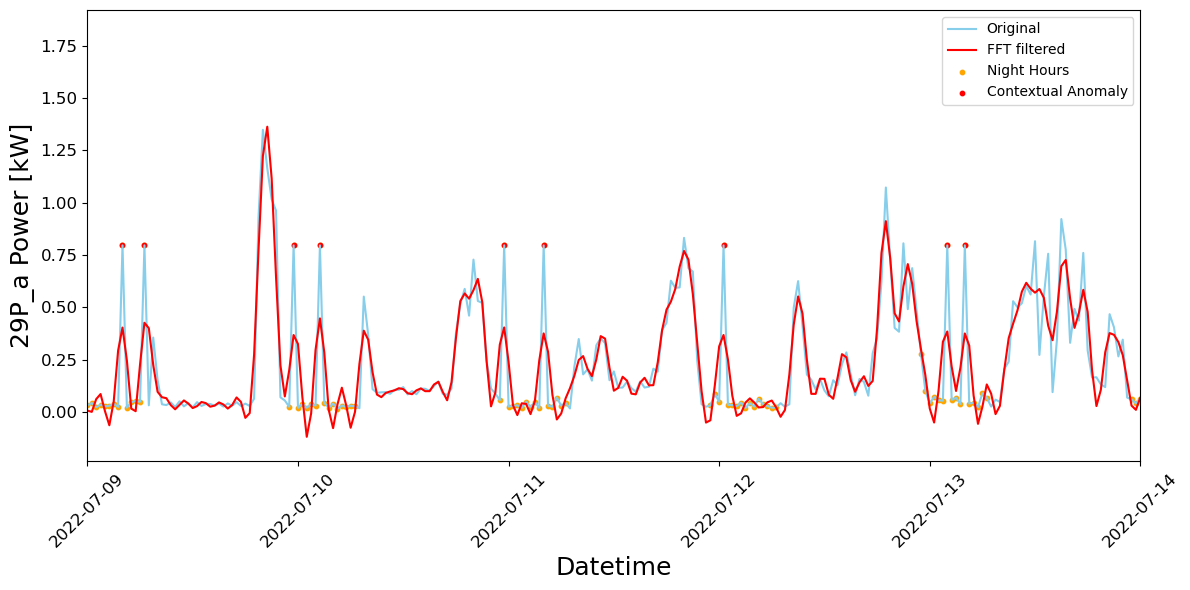} 
    \caption{FFT filter smoothing}
    \label{fig:FFT_smoothing}
\end{figure}

The overall influence of contextual anomalies can be significantly reduced by smoothing with a Fast Fourier Transform filter. The cutoff harmonic for eliminating higher frequencies is chosen visually based on the desired smoothing level. Due to the fact that contextual anomalies represent rapid changes in frequency domain. Although their value is not anomalous itself, after signal reconstruction via inverse Fourier transform, as can be seen in Fig. \ref{fig:FFT_smoothing}, the influence of contextual anomalies is significantly reduced, while the magnitude of non-anomalous data remains practically unchanged, even for peak consumption values. This is a result of filtering the measured consumption signal in frequency domain instead of time domain, as is the conventional approach in anomaly detection. 

\section{Discussion} \label{discussion}

For purposes of comparing results from CS1 and CS2, we decided to develop a conventional data-driven voltage prediction model. Although the same class of methods is applicable in a wide range of data-driven applications. In Tab. \ref{tab:results}, the results from a baseline anomaly-free dataset for the voltage prediction model are compared with those from a dataset with reduced anomaly influence examined in CS1 and CS2. As evident from the metrics, even under extreme conditions, with an even greater impact of anomalies than realistically expected, the model shows a high level of accuracy in predicting voltage magnitudes. Even in the overall worst-case node, the mean absolute error accounts to less than 2\%.

\begin{table}[h]
    \centering
    \caption{Performance metrics for different case studies (per unit)}
    \label{tab:results}
    \resizebox{\columnwidth}{!}{%
    \begin{tabular}{lccc}
        \toprule
        \textbf{Case study} & \textbf{Total MAE} & \textbf{Total MSE} & \textbf{Worst case MAE} \\
        \midrule
        Base case voltage prediction & $2.23 \cdot 10^{-4}$ & $0.1 \cdot 10^{-6}$ & $3.77 \cdot 10^{-4}$ \\
        CS1 & $7.74 \cdot 10^{-4}$ & $1.30 \cdot 10^{-6}$ & $10.60 \cdot 10^{-4}$ \\
        CS2 & $10.39 \cdot 10^{-4}$ & $1.98 \cdot 10^{-6}$ & $15.10 \cdot 10^{-4}$ \\
        \bottomrule
    \end{tabular}}
    
\end{table}

Besides, three case studies presented in Table \ref{tab:results}, \emph{we defined an additional case in which voltage magnitude was forecasted based on the initial dataset in which anomalies were not removed.} This represents a case that is often used in distribution network studies that neglects the possibility of the anomaly occurrence. \emph{If the same model is used for data where anomalies are not filtered, the mean squared error is increased up to $41.51\cdot 10^{-6} p.u.$} The reason behind this is the fact that anomalies will cause significant prediction errors in timesteps where they were recorded.

\begin{figure}[htbp]
    \centering
    \includegraphics[width=0.48\textwidth]{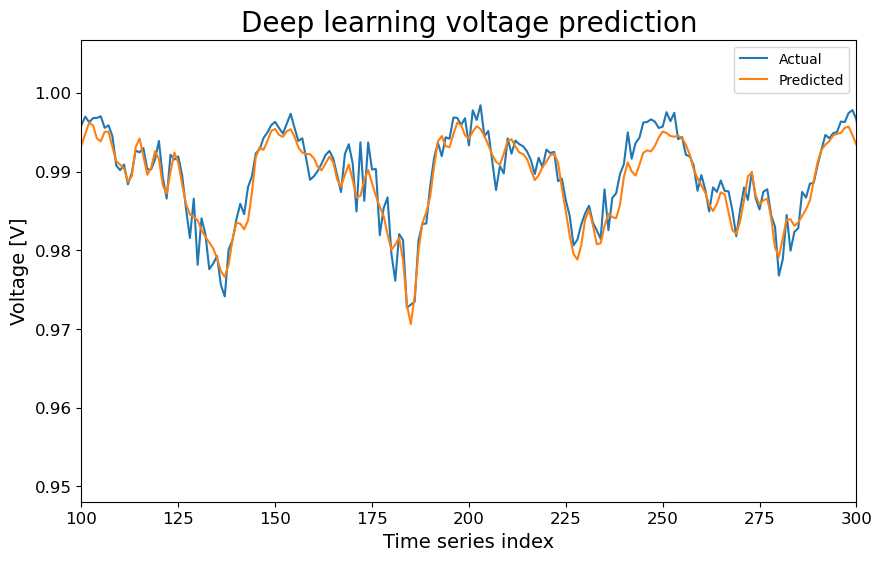} 
    \caption{Voltage prediction for worst case node after FFT filtering}
    \label{fig:voltage_prediction}
\end{figure}

A more detailed visualization of an example of this data-driven voltage prediction is shown in Fig. \ref{fig:voltage_prediction}, where the predicted voltage after the influence of contextual anomalies had been smoothed is plotted against the real voltage magnitudes. 

\section{Conclusion} \label{conclusion}

This paper examined the application of different anomaly detection techniques that can be used in parallel to data-driven and machine learning based models that are sensitive to data quality. 

We have identified two fundamental types of anomalies that frequently appear in smart meter consumption measurement datasets. The first type consists of point anomalies, where a single measurement value deviates from the expected average case. For this type of anomalies, outlier detection methods are used. An isolation forest, a tree-based machine learning algorithm, is trained to detect anomalies by aggregating multiple binary trees, enabling it to isolate anomalies more easily than regular data. By specifying the contamination rate, Isolation forest algorithm eliminates the flagged anomalies and replaces them with a representative value. 

The second identified type are contextual anomalies in time. With this type of anomalies and errors in measurement data, it is not the value of the measurement that deviates from the usual norm, but the timing of the measurement itself. With consumption data, measurements that would be typical during the day and in afternoon hours are completely unusual in the context of the night hours. To reduce the impact of this type of anomalies, filtering in the frequency domain is applied, considering the fact that contextual anomalies are likely to appear as unexpected and rapid spikes, regardless of their magnitude. By applying the Fourier transform, the consumption signal is transformed to the frequency domain, where high-frequency contributions are eliminated using a low-pass filter. The signal is then reconstructed to its representation in the time domain and can be used for further application with reduced and smoothed influence of contextual anomalies in time. Finally, we compared both anomaly detection case studies using the data-driven voltage prediction model as an example. The results show that, with negligible loss of accuracy, it is possible to develop a model that is robust to both point and contextual anomalies. Analyzing the dataset, \emph{we have found that proposed anomaly detection methods work better than conventional statistical detection methods such as Interquartile Range (IQR) method, Chi-squared test or Grubbs' test.} This is especially true for contextual anomalies, where anomalous values would be unrecognized by conventional algorithms that are detecting deviating values.

Future work will include research of different anomaly detection techniques and exploring more subgroups of different anomaly types. Future research will also include comparison between anomaly detection algorithms based on machine learning and statistical anomaly detection, as well as further improvement of the existing models and its evaluation on various real-world datasets.

\bibliography{Literature}

\begin{thebibliography}{10}

\bibitem{8322199}
Y.~Wang, Q.~Chen, T.~Hong, and C.~Kang, ``Review of smart meter data analytics: Applications, methodologies, and challenges,'' {\em IEEE Transactions on Smart Grid}, vol.~10, no.~3, pp.~3125--3148, 2019.

\bibitem{9115822}
X.~Lei, Z.~Yang, J.~Yu, J.~Zhao, Q.~Gao, and H.~Yu, ``Data-driven optimal power flow: A physics-informed machine learning approach,'' {\em IEEE Transactions on Power Systems}, vol.~36, no.~1, pp.~346--354, 2021.

\bibitem{9813554}
K.~Ye, J.~Zhao, N.~Duan, and Y.~Zhang, ``Physics-informed sparse gaussian process for probabilistic stability analysis of large-scale power system with dynamic pvs and loads,'' {\em IEEE Transactions on Power Systems}, vol.~38, no.~3, pp.~2868--2879, 2023.

\bibitem{10005095}
Y.~Liu, S.~Gao, G.~Qiu, T.~Liu, L.~Ding, and J.~Liu, ``A physics-informed action network for transient stability preventive control,'' {\em IEEE Transactions on Power Systems}, vol.~38, no.~2, pp.~1771--1774, 2023.

\bibitem{10517953}
L.~Liu, N.~Shi, D.~Wang, Z.~Ma, Z.~Wang, M.~J. Reno, and J.~A. Azzolini, ``Voltage calculations in secondary distribution networks via physics-inspired neural network using smart meter data,'' {\em IEEE Transactions on Smart Grid}, vol.~15, no.~5, pp.~5205--5218, 2024.

\bibitem{10167837}
B.~Habib, E.~Isufi, W.~van Breda, A.~Jongepier, and J.~L. Cremer, ``Deep statistical solver for distribution system state estimation,'' {\em IEEE Transactions on Power Systems}, vol.~39, no.~2, pp.~4039--4050, 2024.

\bibitem{9975837}
V.~Bassi, L.~F. Ochoa, T.~Alpcan, and C.~Leckie, ``Electrical model-free voltage calculations using neural networks and smart meter data,'' {\em IEEE Transactions on Smart Grid}, vol.~14, no.~4, pp.~3271--3282, 2023.

\bibitem{8813103}
C.~Feng, M.~Sun, and J.~Zhang, ``Reinforced deterministic and probabilistic load forecasting via $q$ -learning dynamic model selection,'' {\em IEEE Transactions on Smart Grid}, vol.~11, no.~2, pp.~1377--1386, 2020.

\bibitem{9528900}
J.~Li, S.~Wei, and W.~Dai, ``Combination of manifold learning and deep learning algorithms for mid-term electrical load forecasting,'' {\em IEEE Transactions on Neural Networks and Learning Systems}, vol.~34, no.~5, pp.~2584--2593, 2023.

\bibitem{9399287}
Y.~Fu, H.~Chai, Z.~Zhen, F.~Wang, X.~Xu, K.~Li, M.~Shafie-Khah, P.~Dehghanian, and J.~P.~S. Catalão, ``Sky image prediction model based on convolutional auto-encoder for minutely solar pv power forecasting,'' {\em IEEE Transactions on Industry Applications}, vol.~57, no.~4, pp.~3272--3281, 2021.

\bibitem{10669038}
G.~Yan and Z.~Li, ``Construction of an outlier-immune data-driven power flow model for model-absent distribution systems,'' {\em IEEE Transactions on Power Systems}, vol.~39, no.~6, pp.~7449--7452, 2024.

\bibitem{8809905}
Y.~Lin and J.~Wang, ``Probabilistic deep autoencoder for power system measurement outlier detection and reconstruction,'' {\em IEEE Transactions on Smart Grid}, vol.~11, no.~2, pp.~1796--1798, 2020.

\bibitem{9760038}
K.~R. Mestav, X.~Wang, and L.~Tong, ``A deep learning approach to anomaly sequence detection for high-resolution monitoring of power systems,'' {\em IEEE Transactions on Power Systems}, vol.~38, no.~1, pp.~4--13, 2023.

\bibitem{ASEFI2023101116}
S.~Asefi, M.~Mitrovic, D.~Ćetenović, V.~Levi, E.~Gryazina, and V.~Terzija, ``Anomaly detection and classification in power system state estimation: Combining model-based and data-driven methods,'' {\em Sustainable Energy, Grids and Networks}, vol.~35, p.~101116, 2023.

\bibitem{9726670}
D.~Singhal, L.~Ahuja, and A.~Seth, ``Anomaly detection in smart meters: Analytical study,'' in {\em 2022 2nd International Conference on Power Electronics \& IoT Applications in Renewable Energy and its Control (PARC)}, pp.~1--5, 2022.

\bibitem{isolation_forest}
F.~T. Liu, K.~Ting, and Z.-H. Zhou, ``Isolation forest,'' pp.~413 -- 422, 01 2009.

\end{thebibliography}
\bibliographystyle{ieeetr}

\vspace{12pt}

\end{document}